\documentclass{jaums}
\theoremstyle{thmit} 
\newtheorem{thm}{Theorem}[section]

\newtheorem{prop}[thm]{Proposition}

\theoremstyle{thmrm} 

\newtheorem*{oldproof}{Proof}
\renewenvironment{proof}[1][{}]{\begin{oldproof}[#1]}{\qed\end{oldproof}}

\usepackage{euscript}

\usepackage{latexsym}
\usepackage{amsmath, amssymb,graphics}
\usepackage[all]{xy}

\usepackage{fullpage}
\renewcommand{\(}{\begin{equation*}}
\renewcommand{\)}{\end{equation*}}
\newcommand{\bea}{\begin{eqnarray*}}
\newcommand{\eea}{\end{eqnarray*}}
\newcommand{\R}{{\mathbb R}}

\newcommand{\Z}{{\mathbb Z}}
\newcommand{\Q}{{\mathbb Q}}

\newcommand{\cL}{\ensuremath{\mathcal L}}

\newcommand{\cT}{\ensuremath{\mathcal T}}
\newcommand{\cW}{\ensuremath{\mathcal W}}

\def\S{\ensuremath{\ES{S}}}
\def\L{\ensuremath{{\cal L}}}
\def\O{\ensuremath{{\cal O}}}

\def\S{\ensuremath{\ES{S}}}
\def\L{\ensuremath{{\cal L}}}

\def\O{\ensuremath{{\cal O}}}

\newcommand{\beq}{\begin{equation}}
\newcommand{\eeq}{\end{equation}}

\newcommand{\ES}[1]{\ensuremath{\EuScript{#1}}}

\numberwithin{equation}{section}

\renewcommand{\(}{\begin{equation}}
\renewcommand{\)}{\end{equation}}

\def\R{{\mathbb R}}
\def\Z{{\mathbb Z}}
\def\Q{{\mathbb Q}}

\def \qed{\mbox{ $\square$}}

\def\1{{\bf 1}}

\def\<{\langle}
\def\>{\rangle}

\def\O{{\cal O}}

\numberwithin{equation}{section}

\renewcommand{\(}{\begin{equation}}
\renewcommand{\)}{\end{equation}}

\begin{document}

\begin{center}
{\Large\bf 
Geometric and topological structures related to M-branes II:
 Twisted String and String${}^c$ structures}
\end{center}
\vspace{1em}

\begin{center}
Hisham Sati 
\footnote{e-mail: {\tt
hsati@math.umd.edu}
}
\end{center}

\begin{center}
Department of Mathematics\\
University of Maryland\\
College Park, MD 20742 
\end{center}

\vspace{0.5cm}
\begin{abstract}
\noindent

The actions, anomalies, and quantization conditions 
allow the M2-brane and the M5-brane to support,
in a natural way, structures beyond Spin on their 
worldvolumes. The main examples are twisted 
String structures. This also extends to 
twisted String${}^c$ structures, which we introduce
and relate to twisted String structures. 
The relation of the C-field to Chern-Simons theory suggests 
the use of the String cobordism category to describe the 
M2-brane.
\end{abstract}

\begin{center}
\it Dedicated to Alan Carey, on the occasion of his 60$\,^{th}$ birthday
\end{center}


\section{Introduction}
In \cite{tcu} we described various geometric and topological structures 
related to the M2-brane (or membrane) and the M5-brane 
(or fivebrane) in M-theory. 
Some of these structures have already been established 
there. Other structures were merely outlined and 
hence deserve more detailed and careful elaboration. In addition,
there are other structures not covered in the above work. This is 
a first in a series of papers which will establish this: expand on structures
eluded to in \cite{tcu} as well as uncover new structures. 
\footnote{As that paper was the starting point of this 
investigation, we will refer to it as part I. Subsequent works will be numbered
accordingly. Thus, this current note will be number II.
} 

\vspace{3mm}
Consider the C-field in M-theory with field strength $G_4$. 
In \cite{SSS3}, the flux quantization condition in M-theory 
on a Spin eleven-manifold $Y^{11}$
\cite{Flux}
\(
G_4 - \frac{1}{2}\lambda = a \in H^4(Y^{11};\Z)
\label{qua}
\)
was recast as defining (essentially) a twisted String structure \cite{Wa}
on $Y^{11}$. 
Here $a$ is the characteristic class of an $E_8$ bundle on $Y^{11}$ and
$\lambda$ is half the first Pontrjagin class $\frac{1}{2}p_1(TY^{11})$ of the
tangent bundle $TY^{11}$ of $Y^{11}$. 
A model for $G_4$ in terms of 
twisted differential cohomology was given in \cite{SSS3}. 
In \cite{tcu} the C-field `potential' $C_3$ was identified as (essentially) 
the String class corresponding to the String structure. 
It is known that the C-field couples electrically to the M2-brane, that 
is the action of the membrane contains a term $\int_{\cW^3} C_3$,
where $\cW^3$ is the membrane worldvolume, an oriented Spin 
three-manifold. The C-field also couples magnetically to the 
M5-brane, that is the fivebrane worldvolume $\cW^6$ will 
couple to $C_6$, the potential corresponding to the Hodge dual 
$*_{11}G_4$, with respect to the metric $g_Y$ on $Y^{11}$. 
Thus, it is natural to consider the questions of existence and 
consequences of String 
structures on the 
worldvolumes $\cW^3$ and  $\cW^6$, rather than just on the 
target spacetime $Y^{11}$. It is the purpose of this note to do just that. 

\vspace{3mm}
The embedding of the worldvolumes of the M-branes in spacetime $Y^{11}$
allows the decomposition of the tangent bundle of $Y^{11}$ into 
tangent bundle of the M-branes and the corresponding normal bundle. 
Almost all the structures we are considering satisfy a two out of three 
principle, that is if two of the bundles above admit a given structure then 
so does the third. Then, for example, if we establish that $Y^{11}$ and the
normal bundle have such structures then so does the M-brane 
worldvolume. However, the situation is not quite as simple, since 
we seeking 
an intrinsic characterization of such structures. That is, we discover them 
via quantization of flux and requiring the partition 
functions to be well-defined, so that we are dealing with 
quantum rather than classical statements, following Witten's work
\cite{Flux} \cite{W-5} \cite{W-Duality}.
Hence, we characterize not only that such structures exist but 
also how they occur.
Furthermore,
``worldvolumes" 
should be interpreted in the appropriate sense as they 
could mean extended worldvolumes, that is 
higher dimensional manifolds obtained from the original worldvolumes
by extending through a circle, taking a bounding manifold, or 
considering disk bundles. The corresponding normal bundles
are modified accordingly.

\begin{thm}
Consider the M-branes as Spin manifolds inside a Spin 
eleven-manifold $Y^{11}$. Then 

\noindent 1. The tangent bundle and the 
normal bundle to the M2-brane each admits 
a twisted String structure. Furthermore, the 
M2-brane worldvolume supports a (differential) String 
cobordism invariant. 

\noindent 2. The (extended) tangent bundle and the 
normal bundle to the M5-brane each admits 
a twisted String structure. 

\end{thm}

\noindent By the extended tangent bundle of a manifold 
we mean the tangent bundle of 
the disk bundle over that manifold.

\vspace{3mm}
In addition to (twisted) String structures, we find that 
String${}^c$ structures, as defined in \cite{CHZ}, appear on 
the worldvolumes. We see that such structures are
closely related to twisted String structures. In addition, we find
that a twisted version of String${}^c$ structures is also
relevant. Thus we are led to define such structures and study some
of their elementary properties. 


\begin{prop} 
Consider a String${}^c$ structure  of a bundle 
$E$ with $\ell$ the Chern class of the line bundle defining the 
Spin${}^c$ structure, and let $Q_1^\alpha (E;\ell)$ denote a
twisted String${}^c$ class, which is $Q_1(E;\ell)- \alpha$,
where $\alpha$ is a degree four integral cocycle. 
 
\noindent {1.} Under change of Spin${}^c$ structure, a String${}^c$ 
structure changes as $Q_1(E;\ell + 2m)=Q_1(E;\ell) - 2 \ell m -2m^2$.

\noindent {2.} Twisted String${}^c$ structures are not quite multiplicative; 
for a fixed $\alpha$
they
satisfy $$Q_1^\alpha(E \oplus E'; \ell + \ell') =
Q_1^\alpha (E;\ell) + Q_1^\alpha (E'; \ell') - \ell \ell'.$$
\end{prop}

\noindent Now we consider the application of twisted String${}^c$ structures to M-branes.

\begin{thm}
Consider the M-branes as Spin${}^c$ manifolds
inside a Spin${}^c$ eleven-manifold $Y^{11}$.
Then the tangent bundle and the 
normal bundle of the M-branes each admits 
a twisted String${}^c$ structure. 


\end{thm}

There are geometric refinements of the String structure discussed
in \cite{R} \cite{Wal} \cite{Bu1} \cite{BN}.  
Differential refinements are discussed in \cite{Bu1} in the 
untwisted case, and in \cite{SSS3} in the twisted case. 
The relation of the C-field to Chern-Simons theory suggests that 
the String cobordism category in which to define the partition function for the M2-brane,
and suggests the use of a String cobordism invariant, providing further
support to ideas presented in \cite{tcu} from a complementary point of view.

\vspace{3mm}
The organization of the paper is very simple. We recall twisted String structures
in section \ref{t St} which we apply to the worldvolumes (and normal bundles)
of the M5-brane and the M2-brane in sections \ref{St M5} and \ref{St M2}, respectively.
We also extend the discussion to the differential case for the M2-brane in 
section \ref{dif}. Next, in section \ref{t Stc}, we consider twisted
String${}^c$ structures, introducing the basic notions in section 
\ref{Def}, and providing the descriptions for the M2-brane and the M5-brane in 
sections \ref{Stc M2} and \ref{Stc M5}, respectively. 
We will use the notation $\lambda$ or $Q_1$ for the first Spin characteristic class,
$Q_1^\alpha$ for the corresponding twisted class, 
and $\lambda^c$ or $Q_1(~~;\ell)$ for the the class in the Spin${}^c$
case.

\section{Twisted String structures}
\label{t St}

\noindent{\bf String structure.}
The first Pontrjagin class $p_1$ for a Spin bundle $E$ is divisible by 
two since $p_1(E) \equiv w_2(E)^2$ mod 2, where 
$w_2(E)$ is the second Stiefel-Whitney class of $E$. 
This then allows the definition of the first Spin characteristic class
$Q_1=\lambda=\frac{1}{2}p_1$, 
which universally is a generator of $H^4({\rm BSpin};\Z)$.
For two vector bundles $E$ and $E'$ admitting Spin structures, 
the first Spin characteristic class is additive (see \cite{Th})
\(
Q_1(E \oplus E')= Q_1(E) \oplus Q_1(E')\;.
\label{Q1 m}
\)
Note that this is an improvement over the corresponding formula
for the Pontrjagin classes (see \cite{MS})
\(
p_1(E \oplus E')=p_1(E) + p_1(E') ~~{\rm mod~} 2{\rm -torsion}\;, 
\)
in the sense that the 2-torsion is automatically taken care of. 
A String structure
on a bundle $E$ or space $X$, originally defined via 
loop spaces \cite{Ki} \cite{CP},
 is a lift of the structure group from 
Spin to String, the 3-connected cover of Spin. That is, the classifying 
map $f: X \to B{\rm Spin}(n)$ of the natural Spin bundle on an $n$-manifold
$X$ is lifted to a map $f': X \to B{\rm String}(n)$ via the fibration 
$K(\Z, 3) \to B{\rm String}(n) \to B{\rm Spin}(n)$. 

\vspace{3mm}
\noindent{\bf Twisted String structure.}
Recall from \cite{Wa} \cite{SSS3} 
that a twisted String structure on a brane $\iota: M \to X$ 
with Spin structure  
classifying map $f: M \to B{\rm Spin}(n)$ is a 
a four-cocycle $\alpha: X \to K(\Z,  4)$ 
and a homotopy $\eta$ between $f^*\lambda=\lambda(M)$ and 
$\iota^* [\alpha]$, as indicated in the diagram
  \(
    \raisebox{20pt}{
    \xymatrix{
       M
       \ar[rr]^f_>{\ }="s"
       \ar[d]_\iota
       &&
       B \mathrm{Spin}(n)
       \ar[d]^{\lambda}
       \\
       X
       \ar[rr]_\alpha^<{\ }="t"
       &&
       K(\Z,4)
       \ar@{=>}^\eta "s"; "t"
    }
    }
    \,. 
\)
Then $M$ has a twisted String structure when 
$\frac{1}{2}p_1(M) + \iota^*[\alpha]=0 \in H^4(M;\Z)$.

\subsection{Twisted String structure and the M5-brane}
\label{St M5}

In \cite{SSS3} the flux
quantization condition in M-theory \eqref{qua} is interpreted essentially 
as an obstruction to the existence of a  twisted String 
structure, and the role of the corresponding higher connection is 
highlighted in \cite{tcu}. The M5-brane six-dimensional worldvolume 
$\cW^6$ admits a map to the target eleven-dimensional 
spacetime $Y^{11}$. The tangent bundle then splits as
$TY^{11}|_{\cW^6}= T\cW^{6} \oplus N\cW^6$, 
where $N\cW^6$ is the corresponding 
normal bundle. 

\vspace{3mm}
\noindent{\bf Flux quantization and twisted String structure on the M5-brane.}
Now we consider topological part of 
the M5-brane worldvolume action. 
Such an action is best described topologically 
via a lift to an eight-dimensional disk bundle over the 
original worldvolume \cite{W-5} \cite{HS},
that is, there is a bundle $\mathbb{D}^2 \to X^8 
\buildrel{\pi^D}\over{\longrightarrow} \cW^6$, described 
as follows.  
 Let $M^7$ be a Spin 7-manifold, which is a circle 
  bundle over $\cW^6$, and which has a C-field $C_3$. 
  Then we have an $E_8$ bundle on $M^7$.
  Let $X^8$ be an oriented 8-manifold with boundary 
  $M^7$ over which $C_3$ extends. 
  This always exists because of the vanishing of the 
  cobordism group $\Omega_7^{\rm Spin}(K(\Z, 4))=0$,
  since $K(\Z,3)\sim E_8$ in this range of dimensions.  
The action includes the term 
\(
S_8=\int_{X^8} G_4 \cup G_4 - G_4 \cup \lambda\;.
\label{m5}
\)
  In \cite{W-5} the action functional 
  was derived
  using the 
  Chern-Simons construction. 
  For $x\in H^4(\cW^6;\Z)$, the construction of the partition 
  function requires defining a $\Z_2$-valued function 
  $\Omega(x)=(-1)^{h(x)}$, where $h(x)$ is an action functional,
  desired to be even, i.e. zero when taken mod 2. 
  Form the 8-dimensional manifold $X^8$ as above, 
  i.e. a disk bundle over $\cW^6$. The element $x$ extends
  to $z= u \cup x$  on $X^8$, where 
  $u \in H^1(S^1;\Z)$.
  For $z=G_4/2\pi$,
  the action is well-defined modulo $2\pi$ and given by 
  \(
  I(C_3)=2\pi \int_{X^8} z\cup z\;.
  \label{cs7}
  \) 
  The Chern-Simons construction requires a division by 
  half, as then the construction will give a line bundle 
  $\cL$ over the intermediate Jacobian $\cT=H^3(\cW^6;\R)/H^3(\cW^6;\Z)$
  so that $c_1(\cL)$ equals the symplectic form $\omega$ 
  on $\cT$, via geometric quantization. 
  
  \vspace{3mm}
  The inability to define $I(C_3)/2$ in general 
  is given by the fact that the intersection form on 
  $H^4(X^8;\Z)$ is not always even. 
  If this were the case then the even-ness of $z^2$ would allow
  the division by 2 and hence give $I(C_3)/2$, well-defined 
  modulo $2\pi$. 
  The mechanism to get around this was proposed in 
  \cite{W-5} as follows. For any $a \in H^4(X^8;\Z)$,
  $a^2 \equiv a \cdot \lambda$ mod 2.
  This is equivalent 
  to the statement that 
  \(
  \frac{1}{8} \int_{X^8} \left( (a- \frac{1}{2}\lambda)^2-\frac{1}{4}\lambda^2 \right)
  ~\in~\Z\;. 
  \)
 This means that the flux quantization condition holds on the 
 eight-dimensional manifold $X^8$, the disk bundle over 
 the worldvolume of the M5-brane,
 $
 \left[\frac{G_4}{2\pi}\right]=\frac{1}{2}\lambda - a$.
   The effect is then to modify the action 
  \eqref{cs7} to \cite{W-5}
  \(
  \widetilde{I}(C_3)=\pi \int_{X^8} \left(z^2- \frac{1}{4}\lambda^2 \right)\;.
  \)
  Since $a$ is integral and $z=\frac{1}{2}\lambda - a$,
    $\frac{1}{2\pi}\widetilde{I}(C_3)$
    is well-defined mod $2\pi$
  and can be used to define the line bundle $\cL$ with 
  $c_1(\cL)=\omega$.

\vspace{3mm}
At the level of the six-dimensional worldvolume $\cW^6$, a similar
condition seems to arise. The dimensional reduction of the action
\eqref{m5} along the disk, i.e. integration over the fiber 
of the two-disk bundle $\pi^D: X^8 \to \cW^6$, gives
\(
S_6=\int_{\cW^6} C_3 \cup h_3 - b_2 \cup \lambda\;,
\label{S6}
\)
where $dC_3=G_4+$ exact terms, $h_3=\pi^{D}_*G_4$.    
Now the variational principle applied to the small 
$B$-field $b_2$ naively gives 
\(
G_4 - \lambda=0 \in H^4(\cW^6;\Z)\;.
\label{sma}
\)
However, we note several ambiguities here. 
First, the action 
\eqref{S6} is not complete, as there are inevitably other terms.
Second, 
there is a subtlety related to the self-duality of the theory
 (see \cite{W-5}).
Third, the process of dimensional reduction on the disk 
assumes $\lambda(X^8)=\lambda(\cW^6)$. Fourth, there
will be contributions to the M5-brane worldvolume from 
$\pi_*(\Theta)$, the integration over the fiber of the 
$S^4$ bundle over $\cW^6$ of the dual $\Theta$ of the
C-field \cite{DFM}. This process of dimensional reduction of the 
disk over $\cW^6$ is mathematically similar to that of 
taking $Y^{11}$ itself to be a disk bundle, and such 
a process involves ambiguities of division by two as
in (the discussion leading to) equation (11.11) in \cite{DFM}.

\vspace{3mm}
  Now consider a modification of the action \eqref{cs7}.
   For $X^8$ Spin, the values of
  the integral
  \(
  h(x)=\int_{X^8} \left( z \cup z + \lambda \cup z \right)
  \label{h of x}
  \)
  is always even. 
  The term $\lambda \cup z$ in \eqref{h of x} means that 
  instead of quantizing 
  the torus
  $\cT$ that parametrizes flat C-fields on $\cW^6$ modulo gauge 
  transformations, the modification implies that 
  one instead is quantizing  another torus $\cT'$,
  which parametrizes, up to gauge transformations, C-fields that
  are no longer flat, but instead have curvature $\frac{1}{2}\lambda$ 
  \cite{W-5} \cite{W-Duality}. The new torus 
  $\cT'$ is isomorphic (not canoncially) 
  to the original torus $\cT$, via 
  the map $C_3 \mapsto C_3 + C_3'$, where 
  $C_3'$ is any 
  C-field of curvature $\frac{1}{2}\lambda$.
  The transformation is 
  \(
  \xymatrix{
  h(x) 
\ar[rr]^{\hspace{-3mm}z\mapsto z-\frac{1}{2}\lambda}
&&
\int_{X^8} z \cup z\;.
}  
\)
  We see that this is simply the shift corresponding to a
  twisted String structure, where $\cT$ corresponds to the cocycle
  (to be viewed as a twist) and $\cT'$ is shifted (by the class that is 
  being twisted by the cocycle).

 \vspace{3mm} 
The above is strengthened, from another angle,
 by Witten's proposal 
\cite{W-Duality}
that $G_4|_{\cW^6}=\theta$, where 
$\theta$ is a torsion class on $\cW^6$, 
something that was verified in \cite{DFM}.  
In this more general case, which includes 
torsion explicitly, we would still have 
a twisted String structure, except that now
the twist is formed out of the original twist and 
this new class $\theta$.
  
  \vspace{3mm}
\noindent{\bf Twisted String structure on the normal bundle of the M5-brane.}
Now we can consider the normal bundle of the M5-brane. 
As indicated in \cite{SSS3}, when the cocycle $\alpha$ represents the 
characteristic class of some bundle $E_2$, 
a twisted String structure on $E_1$ can be viewed as the 
String structure on a difference bundle $E_1 - E_2$. 
Hence, we define the class
\(
Q_1^\alpha(E)=Q_1(E) - \alpha\;.
\)
This class satisfies the following additivity
\(
Q_1^\alpha(E \oplus E')=Q_1^\alpha(E) + Q_1^{\alpha}(E')\;. 
\label{ad q}
\)
This formula can be shown, using \eqref{Q1 m} and additivity of cocycles,
 as follows
\bea
(\lambda + \alpha) (E \oplus E')
&=&
\lambda (E \oplus E') + \alpha (E \oplus E')
\nonumber\\
&=&
\lambda(E) + \lambda (E') + \alpha(E) + \alpha (E')
\nonumber\\
&=&
(\lambda + \alpha) (E) + (\lambda + \alpha) (E')\;.
\eea
Now if we take spacetime $Y^{11}$ to be Spin, then the flux quantization 
condition
\eqref{qua} 
  will give spacetime a twisted 
String structure. The above additivity \eqref{ad q}, applied to the 
split tangent bundle via the embedding of the M5-brane,
will then imply that the normal bundle to the M5-brane will also admit a
twisted String structure.

\subsection{Twisted String structure and the M2-brane}
\label{St M2}

The first Spin characteristic class is multiplicative, as we saw in 
\eqref{Q1 m}. 
This means that, in general, if any two of the three bundles
$E$, $E'$, $E\oplus E'$ are String, then so is the third.
Applying this to the M2-brane we have that if the 
normal bundle admits a Spin structure then so does the 
target space $Y^{11}$. 
This is because, for dimension reasons, the M2-brane worldvolume trivially admits
 a String structure. Nevertheless, there are very interesting 
 consequences of requiring the M2-brane to have a String 
 structure \cite{tcu}. 
 On the other hand, we can  
 consider the same question for the 
 normal bundle of the embedding of $\cW^3$ in $Y^{11}$,
 which is also already considered the same work. Instead, what we 
 do here is consider differential refinements,
 via a discussion which is complementary to that in \cite{tcu}. 
 

\subsubsection{Differential refinement of String structure on the M2-brane}
\label{dif}


Consider the M2-brane with worldvolume 
$\cW^3$, a three-dimensional connected closed Spin 
manifold. Then  $\cW^3$ has a canonical topological 
String structure.    
A topological String structure $\alpha^{\rm top}$ is 
by definition a trivialization of the Spin characteristic 
class $Q_1=\frac{1}{2}p_1(TM) \in H^4(M;\Z)$. 
Since MSpin${}_3=0$, we can find a Spin zero bordism 
$Z^4$ of $\cW^3$. 
An oriented 3-manifold is always Spin. 
In addition to such a manifold admitting a String structure for dimension 
reasons, one can also get a canonical String structure via 
a trivialization of the tangent bundle. 
In fact, the main example used in \cite{Flux} is $S^3$, which is 
parallelizable. 
%
%
The physical significance of a String structure for the M2-brane 
is highlighted in \cite{tcu}.

\vspace{3mm}
A geometric refinement $\alpha$ of $\alpha^{\rm top}$
trivializes the class $Q_1$ at the level of differential forms.
A chosen connection $\nabla^\cW$
on the tangent bundle $T\cW^3$ gives rise to 
a connection on the Spin$(3)$-principal bundle given by the 
Spin structure. Choose an extension $\nabla^Z$ of 
$\nabla^\cW$ from $\cW^3$ to $Z^4$. 
The existence of a connection allows for a geometric 
String structure
\cite{R} \cite{Wal} \cite{Bu1};
a topological String structure $\alpha^{\rm top}$ on $\cW^3$
gives rise to a 3-form $C_\alpha$ which satisfies 
$dC_\alpha= \frac{1}{2}p_1(\nabla^\cW)$.  The Chern-Simons
aspect is described in \cite{SSS1}.

\vspace{3mm}
\noindent{\bf Change of String structure.}
The set of topological String structures on $\cW^3$ is a 
torsor under $H^3(\cW^3;\Z) \cong \Z$. The action of 
$x \in H^3(\cW^3;\Z)$ can  be written as
$
(x, \alpha^{\rm top}) \mapsto \alpha^{top} + x$.
Then 
\(
\int_{\cW^3} C_{\alpha^{\rm top}+x} =
\int_{\cW^3} C_{\alpha^{\rm top}} +
\langle x, [\cW^3]\rangle\;, \quad \forall x \in H^3(\cW^3;\Z)\;,
\label{torsor}
\)
where $\langle x, [\cW^3]\rangle$
is the pairing of the cohomology class $x$ with the fundamental 
homology class $[\cW^3]$ of $\cW^3$.

\vspace{3mm}
\noindent{\bf String bordism invariant.}
We will make use of a String cobordism invariant 
defined in \cite{BN},
\(
d_Z(\cW^3, \alpha):= \frac{1}{2}\int_{Z^4} p_1(\nabla^Z) 
- \int_{\cW^3} C_\alpha\;.
\)
This expression a' priori takes values in $\R$, but turns out 
that \cite{BN}

\noindent 1. $d_Z$ is an integer. 

\noindent 2. Furthermore, it is independent of the choice 
of connections and geometric data of the String structure. 

\noindent 3. The corresponding class 
\(
d(\cW^3, \alpha^{\rm top}):=[d_Z(\cW^3, \alpha^{\rm top}]
\in {\rm MString}_3 \cong \Z_{24}
\)
is a String bordism invariant, so that 
the map
$d: {\rm MString}_3 \to \Z_{24}$ which takes 
$\left[\cW^3, \alpha^{\rm top}\right]$ 
to $d(\cW^3, \alpha^{\rm top})$
is an isomorphism. 

\vspace{3mm}
\noindent From \eqref{torsor}, the invariant for a shifted topological String
structure then takes the form
\(
d_Z(\cW^3, \alpha^{\rm top} +x)=d_Z(\cW^3, \alpha^{\rm top}) -
\langle x, [\cW^3] \rangle\;.
\label{10}
\)

\vspace{3mm}
\noindent{\bf A generator for MString${}_3$.}
Let $\S$ denote the sphere spectrum. There is a unit map
from $\S$ to any other spectrum. Thus let 
$\epsilon: \S \to$ MString be the unit of the ring spectrum 
MString. This is an isomorphism in degree 3,
that is MString${}_3 \cong \S_3 \cong \Z_{24}$
\cite{Ho}.  
The sphere $S^3 \in \R^4$, considered as the boundary of the 
disk $\mathbb{D}^4\in \R^4$, 
has a preferred orientation, Spin structure, and String 
structure $\alpha^{\rm top}$. Let 
${\sf or}_{S^3}\in H^3(S^3;\Z)$ be the orientation class of 
$S^3$.
A generator $g \in$ MString${}_3$
is given in \cite{BN} as
\(
g:=[ S^3, \alpha^{\rm top} - {\sf or}_{S^3}] \in {\rm MString}_3\;.
\)
Then, from \eqref{10}, $d(g)=[1]\in \Z_{24}$, which has order 24 
so that $g \in $ MString${}_3$ is a generator. 

\vspace{3mm}
\noindent{\bf M2-brane and 2-framing.}
 Consider a membrane with worldvolume $\cW^3$,
 a compact connected oriented 3-manifold. 
 At the beginning of this section we considered 
 M2-branes with parallelizable worldvolumes. 
 Now we consider a variation.
 The double of the tangent bundle $2T\cW^3=T\cW^3 \oplus T\cW^3$ 
 has a natural Spin structure arising from uniquely 
 lifting the
 structure group to Spin$(6)$ via the following diagram
 of Lie groups
 \(
 \xymatrix{
 &
 &
 &
 {\rm Spin}(6)
 \ar[d]
 \\
 SO(3) 
 \ar[rr]_{ \hspace{-3mm}\rm diag}
\ar@{..>}[urrr]
 &&
 SO(3)\times SO(3)
\ar[r]
 &
 SO(6)\;.
 }
 \)
 A 2-framing of a closed oriented 3-manifold 
 $\cW^3$ is a Spin-trivialization of the double 
 $2T\cW^3$ of its tangent bundle \cite{At}.
 Let $Z^4$ be an oriented zero-bordism of $\cW^3$. 
Then the 2-framing $\alpha$ at the boundary 
$\partial Z^4\cong \cW^3$ gives rise to a trivialization 
of the Spin bundle $2TZ^4$. This trivialization refines the 
Spin class $\frac{1}{2}p_1(2TZ^4) \in H^4(Z^4;\Z)$ to 
a  relative cohomology class 
$\frac{1}{2}p_1(2TZ^4, \alpha)\in H^4(Z^4, \cW^3;\Z)$. Then 
the quantity 
\(
\sigma(\alpha):= 3 {\rm sign}(Z^4) - 
\left\langle \frac{1}{2}p_1 (2TZ^4, \alpha), [Z^4, \cW^3] 
\right\rangle 
\in \Z
\)
gives an integer parametrized by $\alpha$ and
 does not depend on $Z^4$ \cite{At}. 
 The canonical 2-framing $\alpha_0$ is one for which
 $\sigma(\alpha_0)=0$. 
 
 \vspace{3mm}
 A canonical 2-framing gives rise to a canonical String structure
  \cite{BN}.
 Let $\alpha^{\rm top}$ be any topological String structure on 
 $\cW^3$. The combination 
 \(
 \sigma(\cW^3,, \alpha^{\rm top}):=
 3 {\rm sign}(Z^4) - 2d_Z (\cW^3, \alpha^{\rm top}) \in \Z
 \)
 is independent of the choice of $Z^4$ and has a cohomology class 
 \(
 \sigma (\cW^3):= \left[ 
 3 {\rm sign}(Z^4) - 2d_Z (\cW^3, \alpha^{\rm top})
 \right] =[{\rm sign}(Z^4)] \in \Z_2
 \)
which is also independent of the choice of String structure $\alpha^{\rm top}$.
 Then $\cW^3$ has a unique
  topological String structure $\alpha_0^{\rm top}$ 
  characterized by $\sigma(\cW^3, \alpha_0^{\rm top}) \in \Z_2$.

\vspace{3mm}
\noindent{\bf Eta invariant and an expression intrinsic on $\cW^3$.}
An expression for $d$ which does not depend on the bordism 
$Z^4$ is given in \cite{BN}, which we apply to our situation.
Let $S(\cW^3)$ be the Spin bundle of $\cW^3$. 
Let $V \to \cW^3$ be a real $E_8$ vector bundle with a metric and 
connection, then we can form the
Dirac operator $D_{\cW^3} \otimes V$ which acts on sections of 
the bundle $S(\cW^3) \otimes_\R V$. 
A taming of $D_{\cW^3} \otimes V$ is a self-adjoint operator 
$T$ acting on section of $S(\cW^3) \otimes V$ and given by a smooth 
integral kernel such that $D'=D_{\cW^3} \otimes V+ T$
is invertible. 
The taming is physically a mass term which acts as a regulator in the 
(Pauli-Villars) regularization.
This  modified operator is what is used for the eta-invariant. 
By the Atiyah-Singer index theorem \cite{AS}, the index of $D'$ given by 
\(
{\rm Ind}(Z^4)= -\frac{1}{24}\int_{Z^4}p_1(\nabla^Z) + \eta(\cW^3)\;,
\)
so that the following equality of cohomology classes 
\(
\left[ 
\frac{1}{2}\int_{Z^4} p_1(\nabla^Z)
\right] =[12\eta(\cW^3)]
\)
holds in $\R/\Z$.
Choose a geometric refinement $\alpha$ 
of the topological String structure $\alpha^{\rm top}$
based on the Spin connnection induced by $\nabla^\cW$. Then 
a formula for the String bordism invariant which is intrinsic on 
$\cW^3$ is 
\(
d(\cW^3, \alpha)=
\left[ 12 \eta(\cW^3) - \int_{\cW^3} C_\alpha \right] \in \Z_{24}\;.
\)
This is our proposed (part of the) topological action for the M2-brane
which detects the String structures. 
Such a quantity would a priori appear in the partition function after multiplication
by an integer between 0 and 23. However, the coefficient 1 is favored 
by the M2-brane action, and hence by the partition function. 
See \cite{tcu} for more on the partition function, in which
one should sum over String structures. It is possible that 
the other factors appear when one considers M2-brane multi-instantons. 

\vspace{3mm}
\noindent{\bf The $q$-expansions.}
Consider the series $\theta_W(x,q):=\exp \left[\sum_{k=2}^\infty \frac{2}{(2k)!}
G_{2k}x^{2k}
\right] \in \Q[[q]][[x]]$, where $G_{2k}$ are the Eisenstein series, and 
$\Theta:=K_{\theta_W} \in \Q[[q]][[p_1, p_2, \cdots]]$ is the power series
corresponding to $\theta_W$. In \cite{BN}, 
a String bordism 
 invariant, which involves a $q$-expansion, 
 is defined using $\widetilde{\Phi}:=\Theta \frac{e^{G_2 p_1}-1}{p_1}$.
 For $k=1$ this is given by    
\(
b= \int_{\cW^3} C_\alpha \wedge \widetilde{\Phi}_{[0]}\;,
\)
where $\widetilde{\Phi}_{[0]}=G_2$, the first Einsenstein series, 
which is not a modular form. 
Since $G_2=-\frac{1}{24} + q + \cdots$, the result is in 
$\Z_{24} \oplus \Z[[q]]$. 
Indeed, for $q=0$ this gives $-\frac{1}{24}\int_{\cW^3}C_\alpha$
(cf. \cite{tcu}).

\section{(Twisted) String${}^c$ Structures}
\label{t Stc}

\subsection{Definitions and constructions}
\label{Def}

\noindent{\bf Spin${}^c$ structures in terms of Spin structures.}
There is a nice geometric criterion for the existence of 
a Spin${}^c$ structure (see \cite{LM}). 
Since $U(1)=SO(2)$, there is 
a natural map $f_s: SO(n) \times U(1) \to SO(n+2)$. This
extends, via Whitney sum, to a map of bundles. 
The group
Spin$^c(n)$ can then be defined as the pullback 
by $f_s$ of the covering map ${\rm Spin}(n+2) \to
SO(n+2)$
\(
\xymatrix{
{\rm Spin}^c(n) 
\ar[d]
\ar[rr]
&&
{\rm Spin}(n+2)
\ar[d]
\\
SO(n) \times U(1) 
\ar[rr]^{f_s}
&&
SO(n+2)
}\;.
\)
This implies  that a manifold $M$ is Spin${}^c$, i.e. $TM$ has a Spin${}^c$
structure, iff there is a 
complex line bundle $L$ over $M$ such that 
$TM \oplus L$ has a Spin structure.

\vspace{3mm}
A Spin${}^c$ manifold $M$ has a two-dimensional 
class $c \in H^2(M;\Z)$, which reduces mod 2 to 
$w_2(M)$. On such a manifold $p_1 -c^2$ is divisible 
by 2, and there is an integral characteristic class 
$\lambda$ such that $2\lambda= p_1 - c^2$.
More generally, if $M$ is Spin${}^c$, let 
$J$ be a real two-dimensional vector bundle 
with Euler class $c$ and let $E=TM \oplus J$. 
Then $w_2(E)=0$ and $\lambda(E)$ 
is the corresponding String class for $E$ with 
\(
2\lambda (E) = p_1(E)=p_1(TM) - c^2\;,
\label{p1 c}
\)
and is the String class in the Spin${}^c$ case. This is 
simply the String${}^c$ structure.

\vspace{3mm}
\noindent{\bf Remark on coefficients.}
The quantization condition $G_4 - \frac{1}{2}\lambda \in H^4(Y^{11};\Z)$
also holds for $\lambda$ replaced by any integer multiple of 
$\frac{1}{2}\lambda$, that is for $\frac{1}{2}\lambda$ replaced by 
$\frac{1}{2}(2k+1)\lambda$, for any integer $k$. In this paper
we have chosen $k=0$ as required by the index theorem calculations
to cancel membrane anomalies, as in \cite{Flux}.
Note also that in the discussion leading to \eqref{p1 c},
$p_1 - c^2$ can be replaced by $p_1 - (2m+1)c^2$, where $m$ 
is any integer. Indeed, this is compatible with the discussion in
\cite{CHZ} where $m$ is chosen in a dimension-dependent
way so as to get a generalized Witten genus.  
Since we will not deal with 
modular forms in this note, we will not make such 
distinctions.

\vspace{3mm}
\noindent{\bf (Twisted) String${}^c$ structures in terms of String structures.}
From the above discussion, it seems natural to define a String 
structure corresponding to a Spin${}^c$ structure 
via one corresponding instead to a Spin structure. 
That is, to characterize whether a bundle $E_1$ 
admits a String${}^c$ structure we form another bundle
$E_2=E_1 \oplus L_\R$ over the same space $X$ 
and apply the String construction to $E_2$.
The condition $\lambda(E_2)=0$ then translates to the 
condition $\lambda(E_1) -\frac{1}{2}c^2=0$. 
If we take the first bundle $E_1$ to be the tangent bundle
$TX$ and $E=E_2$, then we form the direct sum 
of bundle via the standard diagonal inclusion. 
Let $TX \buildrel{\pi_T}\over{\longrightarrow}X$
and $L_\R \buildrel{\pi_L}\over{\longrightarrow}X$
be the two indicated vector bundles on $X$. Let
$\Delta: X \to X \times X$ be the diagonal map. 
The Whitney sum $TX \oplus L_\R$ of the
two bundles 
$TX$ and $L_\R$ is the pullback of the Cartesian 
product of $TX$ and $L_\R$ via $\Delta$, 
\(
\xymatrix{
E=TX \oplus L_\R 
\ar[rr] 
\ar[d]
&&
TX \times L_\R
\ar[d]^\pi
\\
X 
\ar[rr]^\Delta 
&&
X \times X
}\;.
\)
Similarly we can provide a definition for the twisted 
String${}^c$ structure. In this case we have a homotopy
between $\lambda(E_2)$ and $\alpha$
 \(
    \xymatrix{
       E_2
       \ar[rr]^f_>>>{\ }="s"
       \ar[drr]_{\alpha}^{\ }="t"
       &&
       B \mathrm{Spin}(n)
       \ar[d]^{\lambda}
       \\
       &&
       K(\mathbb{Z},4)
       \ar@{=>}^\eta "s"; "t"
    }
    \,,
  \)
%
so that $\lambda(E_2) + \alpha =0 \in H^4(X;\Z)$, which 
translates to the condition
\(
\lambda(E_1) + \alpha -\frac{1}{2}c^2=0 \in H^4(X;\Z)\;.
\)
This is the condition for the bundle $E_1$ to admit a twisted 
String${}^c$ structure.

\vspace{3mm}
\noindent{\bf String${}^c$ structures directly.}
We can also give a definition of a String${}^c$ structure 
as a special case of a twisted String structure. Note that, in general,
the latter has a twist given by a degree four integral cocycle, while the former
has a composite cocycle $\frac{1}{2}c^2$, which lives in the wedge of 
$K(\Z,2)$ with itself. There is a map from $K(\Z,2) \wedge K(\Z,2) $ 
to $K(\Z,4)$ given by the cup product. We characterize 
a String${}^c$ structure via the diagram
\(
\xymatrix{
X
\ar[rr]^f_>{\ }="s"
\ar[d]_{c(X)}
\ar[ddrr]^\alpha^<{\ }="t"
&&
B{\rm Spin}^c(n) 
\ar[dd]^{Q_1}
\\
K(\Z,2) 
\ar[d]^{\wedge}="a"
&&
\\
{K(\Z,2) \wedge K(\Z,2)}
\ar[rr]^{\cup}="b"
&&
K(\Z,4)
\ar@{=>}^{\eta_1} "s"+(2,-2); "s"- (8,8)
\ar@{<=}_{\eta_2} "a"+(2,-2) ; "a" +(12,4)
}\;.
\label{C D}
\)
The first homotopy $\eta_1$ gives the relation
$Q_1 + \alpha=0 \in H^4(X;\Z)$ and the second 
homotopy $\eta_2$ gives $\alpha + \frac{1}{2}c^2=0 \in H^4(X;\Z)$.
Combined, the two homotopies then give
\(
Q_1 - \frac{1}{2}c^2=0 \in H^4(X;\Z)\;.
\)
This identifies a String${}^c$ structure as a special case of 
a twisted String structure. 
Note that diagram \eqref{C D} should be modified to account 
for the division of $c^2$ by 2. This is already done in \cite{SSS3}
for the case of twisted String structure, and the current case is 
analogous.


\vspace{3mm}
\noindent{\bf $Q_1$ for a Spin${}^c$ vector bundle.}
Let $E$ be a real vector bundle admitting a Spin${}^c$ structure. This means
that $w_2(E)$ is the reduction mod 2 of an integral class 
$\ell \in H^2(X;\Z)$, i.e. $\rho_2(\ell)=w_2(E)$. 
For $\cL$ a complex line bundle with $c_1(\cL)=\ell$, define
the first Spin${}^c$ characteristic class
\(
Q_1(E;\ell)= Q_1(E - \mathcal{L}) \in H^4(X;\Z)\;.
\)
Then $2Q_1(E;\ell)= p_1(E) - \ell^2$ and the mod 2 reduction 
is $\rho_2(Q_1(E;\ell))=w_4(E)$.

\vspace{3mm}
\noindent{\bf Change in Spin${}^c$ structure.}
Now consider the change in the Spin${}^c$ structure. 
Recall that we defined a twisted String structure
as untwisted String structure of a difference bundle.
We can do the same for a String${}^c$ structure because, 
as we have seen above, 
a String${}^c$ structure can be seen essentially as a specialization of 
a twisted String structure. 
For example, 
take the original line bundle $\cL$ and tensor it with 
a square of a line bundle $L$ of Chern class $c_1(L)=m \in H^2(X;\Z)$, 
so that 
$c_1(\cL \otimes L^2)= \ell + 2m$. Then the Spin class changes as
\bea
Q_1(E; \ell + 2m)&=& Q_1(E - \cL \otimes L^2)
\nonumber\\
&=& Q_1(E;\ell) - 2\ell m - 2m^2\;.
\eea

\vspace{3mm}
\noindent{\bf $Q_1$ for a complex vector bundle.}
Let $E$ be a complex vector bundle. Then $E$ admits 
a Spin structure iff the first Chern class $c_1(E)$ is divisible by 
2, that is $c_1(E)=2n$ for some integer $n$. 
Then the first Spin class is 
\(
Q_1(E)= 2n^2 - c_2(E)\;.
\)
We now consider the String${}^c$ case. 
The twisting by a line bundle can be `untwisted'
in the following sense. 
Noting that $p_1=c_1^2 + 2c_2$,  
if $E$ is a complex vector bundle with $c_1(E)=\ell$ then 
$Q_1(E;\ell)= - c_2(E)$. 


\vspace{3mm}
\noindent{\bf Differential refinement of twisted String${}^c$ 
 structures.} 
 As in the case of twisted String structure,
 a twisted String${}^c$ structure can be refined. 
 The cocycle, the Chern class of the line bundle, and 
 the representative for the String class admit refinements
 as in \cite{SSS3} \cite{Bu1}. Therefore, we get similarly 
 a refinement of the twisted String${}^c$ structure, with 
 expressions similar to those in the twisted String case.
 
  
  \subsection{Twisted String${}^c$ structures 
  and the M2-brane} 
 \label{Stc M2}
 
 M-theory is mostly studied on Spin manifolds. However, one
 can also study the theory on
 Spin${}^c$ manifolds. This has been
 discussed extensively in \cite{DMW-Spinc}.
 In this case there is a global gravitino anomaly
 in the eleven-dimensional supergravity description
 which can be shown to cancel; examples of this
 are considered in \cite{DLP}. Furthermore, since
 $G_4$ couples to the gravitino, there is a 
 correction to the flux quantization which is 
 given in \cite{BEM} in the case of torus bundles. 
 In general, when dealing with M-theory 
 one needs to go beyond the supergravity approximation.
 Hence it is possible that the would-be gravitini 
 need to be replaced by membranes. 
 However, we leave the explicit check to future work.
 
 \vspace{3mm}
 On the other hand, we can consider M2-branes
 with worldvolumes admitting a Spin${}^c$ structure.
 Since $\cW^3$ is a three-dimensional
 compact oriented manifold, 
 it is 
 Spin, and hence also Spin${}^c$. 
 By embedding the M2-brane in spacetime, we get
 a splitting $TY^{11}|_{\cW^3}= T\cW^3 \oplus N \cW^3$. 
 Spin${}^c$ structures satisfy a two out of three principle, so 
 that the normal bundle $N\cW^3$ will be Spin${}^c$. 
  Then the derivation of the flux quantization will be 
  analogous to the Spin case of \cite{Flux}, as was outlined
  in \cite{DMW-Spinc}, and will involve index theory on 
  the normal bundle. The result is 
  \(
  G_4 + \frac{1}{2}\lambda + \frac{1}{4}c_1^2({\sf L}) \in H^4(Y^{11};\Z)\;,
  \label{flux spinc}
  \) 
  where ${\sf L}$ is the complex line bundle associated with the 
  Spin${}^c$ structure (see \cite{DMW-Spinc} for details). 
 
 \vspace{3mm}
 In fact, the way the condition is derived is really from the 
 same condition on the normal bundle together with 
 the triviality of the condition on $\cW^3$. 
 In this setting, we can interpret
  \eqref{flux spinc} as defining a twisted 
 String${}^c$ structure on the normal bundle $N\cW^3$. 
 Therefore, we find twisted 
 String${}^c$ structures both on the M2-brane worldvolume as
 well as on its normal bundle.

  \subsection{Twisted String${}^c$ structures and the
  M5-brane} 
  \label{Stc M5}
  
  Consider the bounding 8-manifold $X^8$ as a Spin${}^c$ manifold
  with a fixed Spin${}^c$ structure. 
  $X^8$ has a two-dimensional class $c \in H^2(X^8;\Z)$
  which reduces mod 2 to $w_2(X^8)$ and which is the 
  Euler class of a 2-dimensional vector bundle $E_2$. 
  Furthermore, $p_1- c^2$ is divisible by 2 so that, as above, 
  there is an integral class $\lambda^c$ (or $Q_1(~~;\ell)$)
  such that $2\lambda^c = p_1 -c^2$. 
  Consider the trivial rank 3 bundle $E_3=X^8 \times \R^3$, and consider 
  the rank five Whiteny 
  sum bundle ${\cal E}_5=E_2 \oplus E_3$ over $X^8$.  
  Consider the unit sphere bundle $S({\cal E}_5)$ over 
  $X^8$ which is a twelve-dimensional Spin manifold 
  $Z^{12}$, and denote the projection by 
  $\pi_{\cal E}: Z^{12}\to X^8$.

  \vspace{3mm}
  Let $x\in H^4(X^8;\Z)$ and $u \in H^4(Z^{12};\Z)$ 
  such that $\pi_*(u)=1$ and $u \cup u=0$.  
 This $u$ can be constructed as the Poincar\'e 
  dual of a section of $\pi$. 
  Now consider an $E_8$ bundle $E$ over $Z^{12}$ 
  with degree four characteristic class 
  $a=u + \pi^*(x)$.
  The Spin${}^c$ characteristic class on $Z^{12}$ is taken to be
  the pullback of the corresponding class on 
  $X^8$, that is $\lambda^c(Z^{12})=\pi^*(\lambda^c(X^8))$.
 The index of the Dirac operator for spinors coupled to 
 the $E_8$ bundle is then \cite{W-Duality}
 \(
 i(E)= \int_{X^8} (x \cup x + \lambda^c(X^8) \cup x)\;. 
 \)
We are now
in a situation similar to that of equation \eqref{h of x},
except that $\lambda$ is replaced by $\lambda^c$. 
Hence, we get
 \(
\frac{1}{2} \lambda^c(X^{10}) + x=0 \in H^4(X^8;\Z)\;,
 \)
 which is a condition for the existence of a
 twisted String${}^c$ structure.
  More properly (and precisely), we seek, as in \cite{W-Duality},
  that $i(E)$ is zero when taken mod 2. A necessary 
  condition to ensure this is to require 
  the twisted String${}^c$ condition. 
  We can again consider the situation on $\cW^6$ rather than on 
  $X^8$. We get
  the same condition if we go through the analysis leading to 
  equation \eqref{sma}.
 
 \vspace{3mm}
\noindent{\bf Geometric refinement.}
 Note that we can get a geometric String structure on the M5--brane. 
 We have done this explicitly for the M2-brane in section
 \ref{dif}, and the extension
 to the M5-brane is somewhat similar. However, there are effects which 
 deserves careful treatment and will be discussed fully elsewhere.

\vspace{3mm}
\noindent{\bf M5-brane and MString.}
 The above discussion at the end of  section \ref{dif} on 2-framing 
 for the M2-brane also makes tantalizing 
 connection to the M5-brane, in a special case. 
To see this, consider the M5-brane with 
 worldvolume $\cW^6=\cW^3 \times \cW^3$. A physically 
 appropriate example is to take
 $\cW^3=S^3$ and $\cW^6=S^3 \times S^3$. 
 Then the trivialization of $2T\cW^3$ can be viewed as a trivialization 
 of $T\cW^6$, by the isomorphism. On the other hand, a trivialization 
 of the Spin bundle gives rise to a canonical topological, and hence 
 geometric, String structure \cite{Bu1}.

  \vspace{5mm}
\noindent {\large \bf Acknowledgements}

\vspace{2mm}
\noindent The author would like to thank Ulrich Bunke, 
Alan Carey, 
Bai-Ling Wang, and 
Weipeng Zhang for useful comments and 
discussions. This paper was written at
 the Max-Planck Institute for Mathematics
in Bonn, whom the author  thanks 
for support and for an inspiring atmosphere.


\end{document}